\documentclass[a4paper,11pt]{article}
\usepackage{pos}
\usepackage{aas_macros}

\title{Astronomical X-ray Polarimetry as a diagnostic for questions of fundamental Physics. What we learned from the Imaging X-Ray Polarimetry Explorer (IXPE).}
\ShortTitle{Fundamental Physics with IXPE}

\author*[a]{Paolo Soffitta}
\author[a]{Enrico Costa}
\onbehalf{on behalf of the IXPE Collaboration}
\affiliation[a]{INAF-IAPS,
  Via Fosso Del Cavaliere 100, 00133, Rome, Italy}

\emailAdd{paolo.soffitta@inaf.it}
\emailAdd{enrico.costa@inaf.it}
\renewcommand{\printHeadAuthors}{Paolo Soffitta and Enrico Costa}
\newcommand{\pscalare}{\mathbin{\vcenter{\hbox{\scalebox{.6}{$\bullet$}}}}}

\abstract{X-ray Astrophysics, which addresses extreme physics in extreme conditions, is particularly well-suited for answering questions related to known physics. Reversely tiny effects, but integrated along sidereal distances, allow to probe extensions of known physics or even new physics. The new window into polarimetry in this energy band, opened by the Imaging X-ray Polarimetry Explorer (IXPE)—a NASA-ASI Small Explorer mission launched on $9^{th}$ December 2021—enables an entirely novel approach, whether used alone or in combination with standard observables such as light curves and spectra and with ddata in other wavelengths. In this paper, we review IXPE’s results after nearly three years of successful operation, focusing on their implications for key questions in Fundamental Physics.}

\FullConference{42nd International Conference on High Energy Physics (ICHEP2024)\\
18-24 July 2024\\
Prague, Czech Republic\\}

\begin{document}
\maketitle 

\section{Introduction}
\label{sec:Intro}
Polarimetry is a branch of X-Ray Astronomy that was proposed but only marginally faced in the early stages of X-Ray Astronomy, but nowadays ha been at last exploited with the Imaging X-ray Polarimetry Explorer (IXPE) a NASA-ASI SMEX mission. 

In the past a certain number of  papers proposed that X-Ray (or $\gamma$-Ray) Polarimetry of celestial sources could provide evidence of phenomena of fundamental physics not proofed with other observables. These predictions were done when a mission of X-Ray Polarimetry was almost a wishful thinking and the expectation on polarization of sources was incomplete and widely conjectural. Nowadays IXPE results are giving a first picture of how most of classes of X-Ray sources perform in terms of this quantity. 
The design and performance of IXPE have been extensively discussed in various publications \cite{Weisskopf2022, Soffitta2021,DiMarco2022,DiMarco2022b,DiMarco2023}. In summary, IXPE features three mirrors, each with an effective area of approximately 200 cm$^2$, and three Detector Units (DUs) separated by an extendable boom, providing a focal length of 4 meters. Each DU contains a Gas Pixel Detector \cite{Baldini2021}. The Detector Units are arranged at 120° intervals to mitigate and control potential systematic effects, and if necessary, to compensate for them. The instrument enables X-ray polarimetry with angular (30'' Half Energy Width), temporal (1-2 $\mu$s accuracy), and energy resolution (1 keV @ 6 keV, Full Width at Half Maximum) in the 2–8 keV energy band, offering sufficient sensitivity to measure polarization even for faint extragalactic sources.

In this paper we  try to verify, for a first time and without any claim of completeness, how the proposed measurement fit with the status of knowledge.
We divide the proposed measurements into two major classes:
\begin{itemize}
    \item Phenomena giving evidence of known physics in extreme conditions
    \item Phenomena giving evidence of new physics
\end{itemize}

\section{Probing Dark Corners of known Physics in Extreme Conditions}
\label{sec:DarkCorners}

\subsection{Vacuum Polarization and Birefringence with Magnetars}
\label{subsec:VacuumPolar}
Magnetars are isolated, rapidly spinning neutron stars with spin periods between 2 and 12 seconds and luminosities in the range of 10$^{35}$ -- 10$^{36}$ erg/s, which are 1 to 2 orders of magnitude higher than their rotational energy losses. This excess luminosity is believed to originate largely from their immense magnetic field energy, which can reach strengths of 10$^{14}$ -- 10$^{15}$ gauss \cite{Thompson1996}. Due to their intense magnetic fields and the presence of polarized vacuum virtual electron-positron pairs, magnetars have been proposed as potential astrophysical laboratories for detecting vacuum polarization effects, which manifest as a non-trivial dielectric tensor $\epsilon$ and permeability tensor $\mu$ (for a review, see \cite{Harding2006,Schubert2000}). These effects can lead to birefringence, causing photons to propagate differently depending on their linear polarization states \cite{Heyl2000} so decoupling
the polarization states of photons travelling within the field. The photon polarization states evolve adiabatically, meaning that the polarization state—ordinary or extraordinary—tracks the changing direction of the magnetic field. At larger distances from the neutron star surface, where the magnetic field strength decreases below a critical value but becomes more uniform (at what is known as the polarization limiting radius), the two polarization states recouple. This re-coupling results in a higher degree of polarization compared to that generated at the surface, where the magnetic field is highly non-uniform.

\begin{figure}[ht!]
    \centering
    \includegraphics[width=0.5\linewidth]{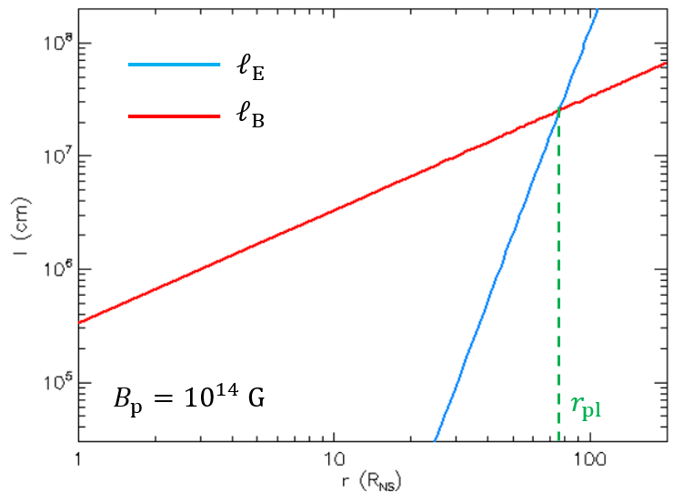}
    \caption{The length scale of the electric-field variation of the radiation coming from the neutron star $\textit{l}_E$}, compared with the length scale of variation of the magnetic field of the magnetosphere of a magnetar $\textit{l}_B$. at relatively smaller distance from the neutron star the electric-field changes on a very short length scale so it can adapt to the local magnetic field, The polarization of the X-ray beam therefore is frozen. At r$_{pl}$, the polarization limiting radius $\textit{l}_E$ becomes larger than $\textit{l}_B$ and the polarization cannot follow anymore the changing magnetic field direction (Figure courtesy of R. Taverna).   
    \label{fig:Polarization limiting radius}
\end{figure}
IXPE observed five Magnetars so far. Results on four of them were published (4U 0162+61 \cite{Taverna2022},1RXS J170849.0-400910 \cite{Zane2023}, SGR 1806-20 \cite{Turolla2023}, 1E 2259+586\cite{Heyl2024}). 
Each of the four magnetars exhibited distinctly different behavior in polarization degree and angle as a function of energy, which has been modeled as arising from various possible emitting regions on the neutron star's surface. These regions include condensed and gaseous atmospheres located on relatively small areas of the neutron star. Additionally, resonant Compton scattering of photons from the neutron star’s surface onto electron-positron pairs flowing through the twisted magnetic field has been proposed to account for the higher polarization degree observed in 4U 0162+61 and the 90$^\circ$ position angle shift.

In emission modeling, quantum electrodynamics (QED) effects, such as vacuum birefringence, were considered but were not strictly necessary. The small emitting regions alone could account for the observed high polarization degree. Nevertheless, the accurate modeling of the phase-resolved polarization angle behavior of 4U 0162+61 using the Rotating Vector Model suggests the possible presence of this non-linear QED effects. This is because the model implies a highly dipolar magnetic field that is only achievable at larger distances from the neutron star, rather than at the surface.

Recently, Lai \cite{Lai2023} proposed that the polarimetric results from IXPE could be entirely of magnetospheric origin. The magnetosphere may be mildly ion-filled and display signatures of QED vacuum resonance, where plasma and vacuum effects counterbalance, resulting in a switch in the polarization angle.
In fact, it is widely recognized that proving the necessary presence of vacuum polarization requires observing a very high polarization degree alongside a very small pulsed fraction in the light curve. The latter indicates a large emitting region, which would imply a small polarization degree at the neutron star’s 
surface with this unique non linear QED responsible for the large X-ray polarimetry.

\subsection{General Relativity in Strong Fields}
\label{subsec:GeneralRelativity}
In the weak-field regime, General Relativity \cite{Einstein1916} is tested against different gravitational theories \cite{Will2018} through direct measurements. For example, nearby artificial satellites are used to study the motion of “test” particles along geodesics of spacetime \cite{Lucchesi2010,Lucchesi2014,Lucchesi2020,Lucchesi2021,Lucchesi2024} and \cite{Ciufolini2004,Ciufolini2019} or by testing the equivalence principle by satellites \cite{Nobili2012,Toumboul2022}. In X-ray astronomy, we rely on the influence of strong gravitational fields around black holes and neutron stars. These fields affect the geodesics of photons from nearby emission sites of the compact object toward the observer and cause gravitational redshifts that distort the spectral features and spectral slopes of photons  (see for a review \cite{Fabian1989,Reynolds2021}). 
Indeed, X-ray polarization has long been suggested as an indication of strong gravity acting in the vicinity of black hole systems \cite{Connors1977,Stark1977,Connors1980}. This effect manifests as a rotation of the polarization vector with energy due to the peculiar geodesics of spacetime. In all these works, General Relativity paved the way to determine the spin of the black hole, since the faster the spin, the larger the rotation of the polarization angle with energy. This is a particularly compelling argument, as different methods—such as reflection \cite{Reis2009}, continuum fitting \cite{Shafee2006}, and quasi-periodic oscillations (QPOs) \cite{Abramowicz2001}—provide different results, at least for GRO J1655-40.
IXPE observations of black hole binaries in the high-soft state—when emission is thermally dominated by a multi-temperature blackbody including relativistic effects \cite{Novikov1973}—have so far detected positive polarization from 4U 1957+11 \cite{Marra2024} and Cyg X-1 \cite{Steiner2024}. In these cases, the energy-dependent (increasing with energy) polarization degree and the polarization angle favor high spin ($a_{*} > 0.96$), with $a_{*}$ being the black hole's angular momentum normalized to its maximum value. This is in agreement with most of the results obtained so far with the continuum fitting method (\cite{Barillier2023} for 4U 1957+11 and \cite{Zdziarski2024} for Cyg X-1).
The IXPE observation of LMC X-3 \cite{Svoboda2024} suggests a low-spin value ($\sim 0.2$), as indicated by \cite{Steiner2014}. In all these observations, a relativistic standard disk model \cite{Novikov1973} fits the data. However, while for 4U 1957+11 and Cyg X-1, X-ray polarimetry requires radiation returning to the disk \cite{Schnittman2010}—and then being reflected off the disk to explain the large and energy-increasing polarization degree—this is not required for LMC X-3, as the disk's inner radius is far from the Innermost Stable Circular Orbit.
For 4U 1630-47 \cite{Ratheesh2023}, the IXPE data tell a completely different story. The observations require a more complicated geometrically thick disk model with mildly relativistic (0.5 c) outflows and a low to intermediate spin (a$_{*}$ $\leq$ 0.7).

\subsection{Equation of State of Neutron Stars}
\label{subsec:EquatState}
Neutron stars, such as radio pulsars and, in X-ray astronomy, millisecond binary pulsars, serve as ideal astrophysical laboratories for testing and constraining the equation of state in the cold, dense regime of matter. In these sources, the magnetic field is typically weak enough that it does not influence the emission pattern. Consequently, the emission depends primarily on the gravitational field at the surface of the compact object (for a review of astrophysical observations related to the determination of the equation of state, see \cite{Poutanen2020}). Among the various methods proposed for deriving the mass-to-radius ratio of neutron stars, photospheric radius expansion for accreting millisecond pulsars (see \cite{Nattila2022} for a review) and precise pulse profile measurements for rotation/accreting powered millisecond pulsars have been suggested \cite{Watts2016,Miller2019,Nattila2022}.
However, the determination of the mass-to-radius ratio from pulse profiles can be degenerate in terms of geometrical parameters, such as the angle between the dipole axis and the rotation axis, as well as the orientation of the rotation axis itself (\cite{Viironen2004}). X-ray polarimetry, by resolving this degeneracy using Rotating Vector Model fitting to the data, could potentially help to better constrain the fiducial region and refine the results obtained from timing measurements \cite{Poutanen2020}.

As a matter of fact, IXPE has so far observed only one  millisecond (accretion-powered) binary pulsar in outburst as a target of opportunity \cite{Papitto2024}. The higher luminosity of accretion-powered millisecond pulsars compared to rotation-powered ones allows for shorter observing times. However, despite trying to constrain the geometrical parameters by applying the rotating vector model to the IXPE data, the small pulsed fraction of U and Q phase resolved light-curve, almost indistinguishable  from a constant, makes impossible their derivation to obtain an accurate equation of state, yet.

In the future, observing additional accreting millisecond pulsars—potentially as IXPE Targets of Opportunity—alongside NICER will be essential to obtain robust constraints on the mass-to-radius ratio of neutron stars.

\section{New Physics}
\label{sec:New Physics}

\subsection{Quantum Gravity}
\label{sec:QuantumGravity}
Theories of Quantum Gravity foresee that Lorentz invariance may be violated at the Planck scale, 

\begin{equation}
    E_P=\sqrt{\dfrac{\hbar c^5}{G}}=1.22 \times10^{19} GeV
\end{equation}

Of course all the energies observed in  reality with our experiments are very far from these values and these effects must be strongly suppressed. With photons propagating over astrophysical distances  these tiny effects will accumulate along the path and this can enable some of the most realistic observational tests. 

In 1999 Gambini and Pullin\cite{Gambini1999} in the frame of a theory belonging to the family of Loop Quantum Gravity proposed that at very high energies the two modes of circular polarization would deviate from c of a quantity increasing with the energy, eventually producing a rotation of the linear polarization plane proportional to distance and to a certain power of the energy, with $E^2$ as the baseline.

On december 12 2002 RHESSI detected a strong GRB. The team claimed for a very high level of polarization\cite{CoburnBoggs2003} in the 0.15 - 2.0 KeV band. Igor Mitrofanov\cite{Mitrofanov2003} submitted a letter with the argument that when a measurement cover such a large energy interval the effect predicted by Gambini and Pullin would \textit{shuffle} the photon polarization and a strong polarization would exclude the effect or better set a strong limit to the coupling constant. Soon after Philip Kaaret\cite{Kaaret2004} submitted another letter  deriving a much more conservative but much more robust limit from the comparison of polarization angle of the Crab between 2.6 and 5.2 keV, the two window of OSO-8. These are the prototypes of inference on QG from gamma or X ray data. After the claim of gamma-ray polarization of the Crab with INTEGRAL on 2008\cite{Dean2008} a new UL was proposed. With the claim of detections of high polarization in GRBs the rush to the lowest upper limit arrived to GRB041219A \cite{Laurent2011}. Of course given a dependence on $E^2$, measurements in the X-Ray band, although providing a sensitive measurement of angles, cannot compete with a gamma-ray measurement. Unfortunately most of these Polarization detections are achieved with instruments not designed and not calibrated as polarimeters. Recent data from the sensitive and well calibrated POLAR mission\cite{Kole2020} suggest that GRBs polarization is typically below 20$\%$, contrary to what found by previous missions. Deriving inference on new physics requires a better assessed picture. For the higher energies we expect that the COSI mission\cite{Tomsick2023} will provide a sound set of data.

Another effort to derive observable birefringence as a low energy effect of more fundamental high energy theory, is based on the Standard-Model Extension1,2 (SME). If Lorentz symmetry is broken, the phase velocity of light in vacuum may depend on the photon energy, polarization, and direction of propagation. The SME is an effective field theory framework that extends the Standard Model of particle physics by introducing new, terms in the Lagrangian violating Lorentz, and CPT, while conserving the charge, energy, and momentum. It allows a categorization of small energy effects, described in essence by a set of coefficients that are characterized in part by the mass dimension d of the corresponding operator. 
 The effect on polarization can be expressed with a modified equation for vacuum dispersion:

\begin{equation}
    E \approx p (1-\varsigma^{(0)}\pm \sqrt{(\varsigma^{(1)})^2+(\varsigma^{(2)})^2+(\varsigma^{(3)})^2})
\end{equation}

The term $\varsigma=(\varsigma^{1(d)},\varsigma^{2(d)},\varsigma^{3(d)})$ is the so-called birefringence axis\cite{Kostelecky2013,Friedman2020}. The coefficients can be expanded in spherical harmonics and in the mass dimension d. The expansion gives coefficients that express the deviation from Lorentz Invariance. In practice from any astrophysical observation a value for some of the parameters can be derived and his can be interpreted in the frame of different theories \cite{Kislat2019,Gerasimov2021}. A major difference is that between odd and even CPT violations. The effect of LIV on polarizaton can be be expressed in terms of Stokes vector \textbf{s }.
and its change in time during the propagation.
\begin{equation} 
    \dfrac{d\textbf{s}}{dt}=2E\varsigma \times\textbf{s}\  with\ \varsigma=(\varsigma^{1})
\end{equation}
In the CPT odd violation circular polarization is preserved. With CPT even violation linear and circular polarization mix\cite{Kislat2019}. By the study of sources at large red-shift limits to the parameters can be derived. From the analysis of circular and linear polarization in the optical band of AGNs and GRB afterglows limits for two parameters have been derived\cite{Kislat2017,Kislat2018, Gerasimov2021}. Recently Kislat has performed \cite{Kislat2023} a very early analysis of the measurement of linear polarization of three extragalactic sources by IXPE (MK421, MK501 and Circinus Galaxy)and combined with already exploited optical data. With respect to optical data only the analysis improves the anisotropic limits of 4 orders of magnitude and the isotropic one of one order of magnitude.

\subsection{Axions \& Axion-like Particles}
\label{subsec:Axions&ALPD}
Axion is the boson proposed by Peccei and Quinn as a natural solution to the strong CP problem. The axion interacts with
fermions, two gluons, and two photons (e.g. $\mathcal{L}$$\sim$ $g_{a\gamma}$ \textbf{B}$\pscalare$\textbf{E}). The Axion mass and the two photon coupling constant are unknown but they are linearly related \cite{Ayala2014}: 

\begin{equation}
    g_{a\gamma} = 2 \times GeV^{-1}\zeta(\frac{m_a}{1eV})
\end{equation}

\begin{figure}
    \centering
    \includegraphics[width=0.5\linewidth]{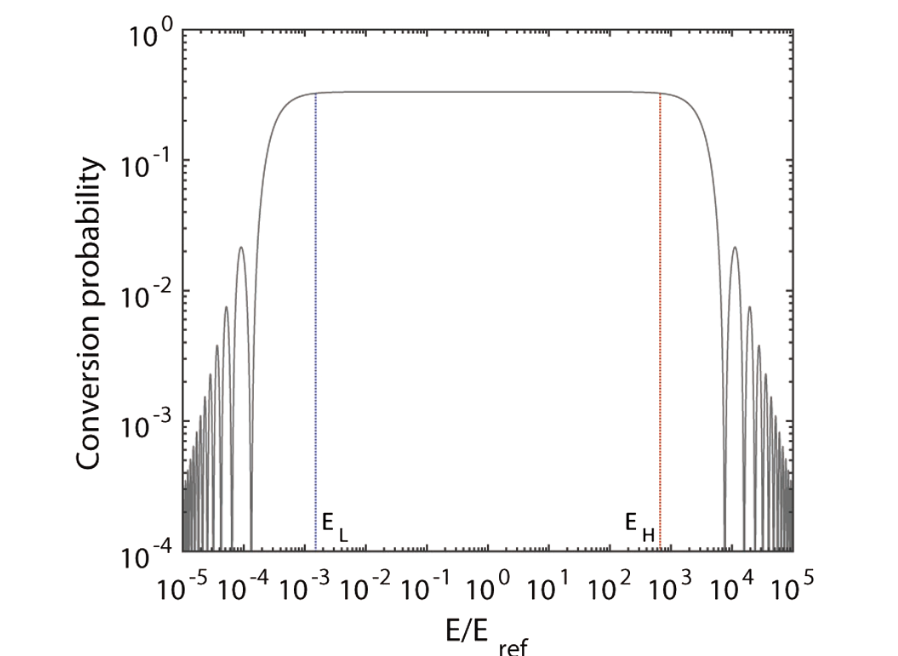}
    \caption{The region below $E_L$ and above $E_H$ are those of Weak Coupling. The region in between is that of Strong Coupling.}
    \label{fig:enter-coupling}
\end{figure}

Axion Like Particles (ALP) are predicted by many models extending Standard Model in particular from Superstring and Brane. They resemble Axions but are different in the sense that do not couple with gluons and fermions and the coupling constant and the mass are not related.
They are a strong candidate for dark matter. The quest to proof their existence and to fix the main features is performed by numerous ground  based experiments or  analysis of  astrophysical data. Within the astrophysical proofs polarimetry is always a good candidate because the ALP - photon oscillation is sensitive to the direction of magnetic field. The oscillation in presence of magnetic field changes the polarization property of a photon beam.
\begin{itemize}
    \item Linearly polarized X-rays passing through magnetic field reduce their parallel component while perpendicular component is unchanged. This can produce a rotation of the plane of polarization of a polarized beam or the onset of a polarization on an unpolarized beam
    \item Further vacuum birefringence may change linear to
elliptical polarization.
\end{itemize}
The probability of ALP-Photon conversion can be subdivided in three zones\cite{Galanti2018} as shown in \ref{fig:enter-coupling}
In the regions of weak mixing ($E\leq E_L$ and $E\geq E_H$) the probability has high oscillations with energy.  The region of major probability in the IXPE band is that of Low Energy weak mixing. The boundaries depend on various parameters E, $m_a$, $g_{a\gamma\gamma}$, plus the magnetic field  B and the plasma frequency  $\omega _L$, that specific of the environment and and are not known \textit{a priori} so that the evidence for the effect can be searched on both spectrum and polarization (and in fact has been already searched with Chandra grating spectra). 

As in other issues of this search we can identify two guidelines:
\begin{itemize}
    \item Search how the ALP-photon coupling modifies polarization properties of some class of sources.
    \item Search polarization introduced on sources that can be reasonably assumed to be not polarized at the origin
\end{itemize}

Of course the second approach is much more robust but not necessarily such a class of sources exist with the needed flux to perform a sensitive measurement.\\
In the past it was proposed that if the intergalactic fields ar organized in domains with magnetic fields aligned on a few MPc scale, the photons polarized at the source could be depolarized at the observer. Conversely  a significant polarization could be detected  from an originally unpolarized source. The computation (at energies higher than those of IXPE and in regime of Strong Mixing) was applied to Gamma Ray Burst Polarization with a certain hypothesis on the starting distribution\cite{Bassan2010}.  But, as we already mentioned, the polarization of GRBs is still  a very open issue.
More recently, thanks to the first results of IXPE and in view of future data at higher energies, especially by COSI mission, Galanti and Rocadelli computed the effect of ALP-photon oscillations on Blazars, as the best class of sources highly polarized at large distance,  and on Galaxy Clusters, which, in absence of bright AGN inside, are likely the less polarized source of X-Rays\cite{Galanti2023b,Galanti2023c}. They added a detailed computation of the effect of magnetic fields in the environment of the source and in our galaxy. These effects come out to be comparable or larger than the effects of intergalactic fields, but at least the magnetic fields in our galaxy are well mapped.

The results on the Blazars could be significant, IXPE has observed $\simeq 14$ blazars. A few have been observed more than once. The difference from object to object and the variability of the same object suggest that disentangling the effect of ALP from all other effects is something that can be the end point of a long path following the observation of many more objects.

\begin{figure}
    \centering
    \includegraphics[width=0.5\linewidth]{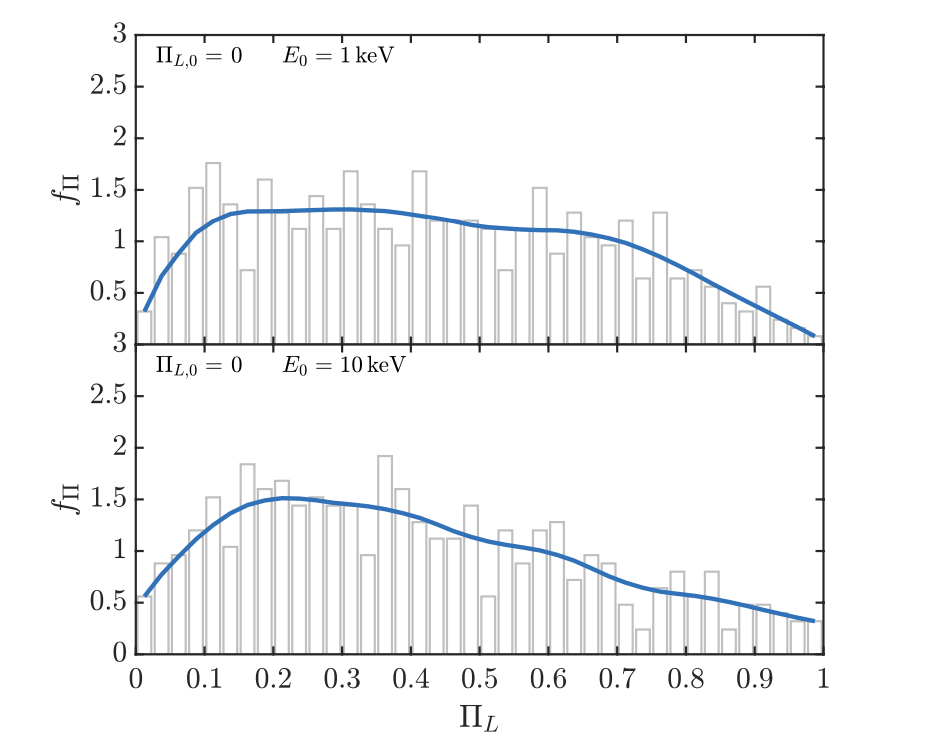}
    \caption{Perseus Cluster. Probability density function for the final degree of linear polarization $\Pi_L$ at 1 keV (upper panel) and 10 keV (lower panel)\cite{Galanti2023c}}
    \label{fig:enter-PerseusProb}
\end{figure}

On the contrary the results on clusters are much more attractive. The X-Rays of clusters, by spectral data, are thermal and should have a polarization that has been predicted\cite{Komarov2016} to be lower that 0.1\%. Computations for Perseus and Coma cluster predict a high probability to find higher values. In \ref{fig:enter-PerseusProb} we show the probability density of final polarization starting from an initial polarization of 0. The parameters of the simulation are $g_{a\gamma\gamma}$ = 0.5 $\times$ 10$^{-11}$ GeV$^{-1}$, $m_a$$\leq$ 10$^{-11}$ eV.
Computations for Coma Cluster give numbers of the same order. IXPE with a long pointing can detect polarizations above 4\% on either Perseus or Coma. Also other calculations arrive to optimistic results, although tuned to energies slightly higher than those of IXPE\cite{Day2018}.
This is likely one of the most promising use of IXPE for a measure of Fundamental Physics.

\acknowledgments
The Imaging X-ray Polarimetry Explorer (IXPE) is a joint US and Italian mission.  The US contribution is supported by the National Aeronautics and Space Administration (NASA) and led and managed by its Marshall Space Flight Center (MSFC), with industry partner Ball Aerospace (now, BAE Systems).  The Italian contribution is supported by the Italian Space Agency (Agenzia Spaziale Italiana, ASI) through contract ASI-OHBI-2022-13-I.0, agreements ASI-INAF-2022-19-HH.0 and ASI-INFN-2017.13-H0, and its Space Science Data Center (SSDC) with agreements ASI-INAF-2022-14-HH.0 and ASI-INFN 2021-43-HH.0, and by the Istituto Nazionale di Astrofisica (INAF) and the Istituto Nazionale di Fisica Nucleare (INFN) in Italy.  This research used data products provided by the IXPE Team (MSFC, SSDC, INAF, and INFN) and distributed with additional software tools by the High-Energy Astrophysics Science Archive Research Center (HEASARC), at NASA Goddard Space Flight Center (GSFC). 

The authors acknowledge useful discussions with  Alessandro Di Marco (INAF-IAPS), Fabian Kislat (University of New Hampshire), David Lucchesi
(INAF-IAPS), Alessandro Papitto (INAF-OAR), Roberto Taverna (Padua University)

\bibliographystyle{JHEP}

\bibliography{References}


\end{document}